\documentclass[twocolumn,seceq]{jpsj3} 

\usepackage{graphicx}
\usepackage{dcolumn}
\usepackage{bm}
\usepackage{graphics,color}

\title{Lattice Distortion to Stabilize the Spin-Peierls State in CeRu$_{\textbf{2}}$Al$_{\textbf{10}}$} 

\author{Katsurou \textsc{Hanzawa}}
\inst{%
Department of Physics, Faculty of Science and Technology,
Tokyo University of Science, Noda 278-8510, Japan 
}%

\recdate{\today}

\abst{
We propose a relevant lattice distortion to stabilize the spin-Peierls state in CeRu$_2$Al$_{10}$,
compatible with the $^{27}$Al-NQR spectra and the neutron diffraction patterns on the assumption
that Al atoms at Al(1) to Al(4) sites displace toward their respective neighboring Ce ions below $T_0$
$=27$~K, leaving Ce, Ru, and Al at Al(5) immobile.
Due to a resulting regular lattice distortion, 
the assembly of one-dimensional spin-Peierls order on each zigzag chain along the $c$-axis predicted by the author becomes
a three-dimensional order with $\bm{Q} = (0, 0, 1)$, where bonds connecting dimerized pair of Ce ions form an \textit{antiferro} ordering 
in a body-centered structure.
}

\kword{spin-Peierls transition, dimerization, lattice distortion, CeRu$_{\textsf{2}}$Al$_{\textsf{10}}$, NQR, Neutron diffraction}

\begin{document}
\maketitle

\section{Introduction}

The novel phase transition observed at $T_0=$ 27~K in CeRu$_2$Al$_{10}$\cite{Strydom,Nishioka,Matsumura,Tanida}
has been investigated theoretically by the present author\cite{Hanzawa} as a spin-Peierls transition on one-dimensional (1D) zigzag chains
formed by nearest neighbor Ce ions along the $c$-axis.
Recently, 
Robert~\textit{et al.}\cite{Robert} have 
reported an observation of magnetic excitation at an energy of 8 meV ($\approx$ 90 K)
below $T_0$
in inelastic neutron scattering experiments for CeRu$_2$Al$_{10}$, which may confirm our proposal of the spin-Peierls transition.
Robert~\textit{et al.}\cite{Robert} have also shown the appearance of forbidden reflections such as 101 below $T_0$ in powder diffraction patterns, suggesting
a breaking or relaxing of the invariance with respect to the centering translation of (1/2, 1/2, 0).
Here, we propose a relevant lattice distortion to stabilize the spin-Peierls state compatible with the neutron diffraction patterns\cite{Robert} and
the $^{27}$Al-NQR spectra,\cite{Matsumura} 
in which four $^{27}$Al-NQR signals assigned to Al(1) to Al(4) sites respectively split into two peaks below $T_0$, while that to Al(5) site does not.
We consider displacements of light Al ions toward their respective neighboring Ce ions on the assumption that
heavy Ru and Ce do not displace below $T_0$.
Among five distinctive Al sites, only Al(5) sites possesses two equivalent neighboring Ce sites, and hence cannot displace in the spin-Peierls state.


CeRu$_2$Al$_{10}$ crystallizes in the orthorhombic YbFe$_2$Al$_{10}$ type structure ($Cmcm$, \#63)\cite{Hahn}
with lattice constants $a=$ 9.1272 \AA, $b=$ 10.282 \AA, and $c=$ 9.1902 \AA,\cite{Thiede,Tursina}
whose unit cell is shown in Fig.~1, where Ce, Ru, and Al atoms are represented by large black, middle white, and small colored spheres, respectively.
There are five distinctive Al sites: Al(1) in 8g site (orange spheres), Al(2) in 8g (yellow),  Al(3) in 8f (blue), Al(4) in 8f (green), and Al(5) in 8e (red).
It should be noted that the unit cell of Fig.~1 is conventional and the primitive cell is half of it, because of
the base-centered orthorhombic structure 
possessing the centering translation of ($a$/2, $b$/2, 0).
In Fig.~1, four Ce sites in 4c are numbered from [1] to [4], and the cyan bonds represent
the dimerization of Ce ions in the spin-Peierls state proposed by the author.\cite{Hanzawa}
The ordering will become to be three dimensional as shown in Fig.~1,
with the ordering vector $\bm{Q} = (0, 0, 1)2\pi/c$,
due to regular lattice displacements predicted below.

\begin{figure}[tb]
\begin{center}
\scalebox{.5}{\includegraphics{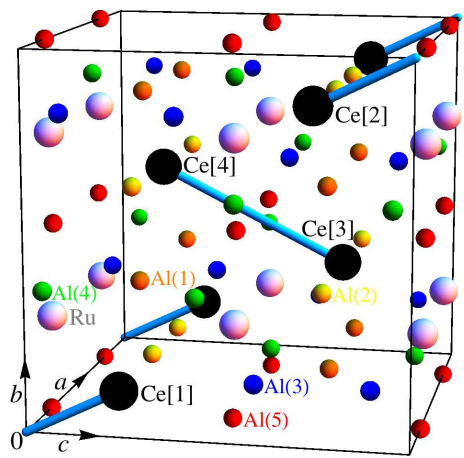}}
\end{center}
\caption{ 
Crystal structure of CeRu$_2$Al$_{10}$ ($Cmcm, Z=4$) in the framework of unit cell, with Ce (large black spheres), 
Ru (middle white spheres), and Al (small colored spheres): Al(1) (orange), Al(2) (yellow), Al(3) (blue), Al(4) (green), and Al(5) (red).
Cyan bonds represent the dimerization of Ce ions in the spin-Peierls state.
}
\label{f1}
\end{figure}

Figure 2 shows the atomic environment of Ce, surrounded by neighboring 20 atoms of 4 Ru, 4 Al(1), 2 Al(2), 4 Al(3), 2 Al(4), and 4 Al(5).
This figure is illustrated for Ce[3], 
which is the same as for Ce[2], while that for Ce[1], as well as for Ce[4], is simply obtained by turning the upside of $b$-axis down.
As shown in Fig.~2, 4 Al(1) and 2 Al(2) atoms, connected by orange lines, surround Ce in an $ab$ plane, whereas
4 Al(3) and 2 Al(4), connected by blue lines, surround Ce in a $bc$ plane.
Note that 4 Ru, 2 Al(2), 2 Al(4), and 4 Al(5) atoms surround Ce equivalently with distances of
3.488 \AA (Ru), 3.203 \AA (Al(2)), 3.188 \AA (Al(4)), 3.349 \AA (Al(5)).
On the other hand,
2 of 4 Al(1) are located at nearest neighbor (n.n.) sites to Ce separated by 3.212 \AA, while other 2 are at next nearest neighbor (n.n.n.) sites to Ce separated by 3.666 \AA;
similarly, 2 of 4 Al(3) at n.n. sites by 3.230 \AA, while other 2 at n.n.n. sites by 3.249 \AA.
These n.n.n. sites of 2 Al(1) and 2 Al(3) are n.n. sites for other Ce sites.
It should be noted that each Al(1), Al(2), Al(3), and Al(4) site has only one n.n. Ce site, respectively,
whereas Al(5), and also Ru, has 2 equivalent n.n. Ce sites.
This uniqueness of Al(5) site among Al sites will be decisive in consideration of possible lattice distortion in the following.

\begin{figure}[tb]
\begin{center}
\scalebox{.7}{\includegraphics{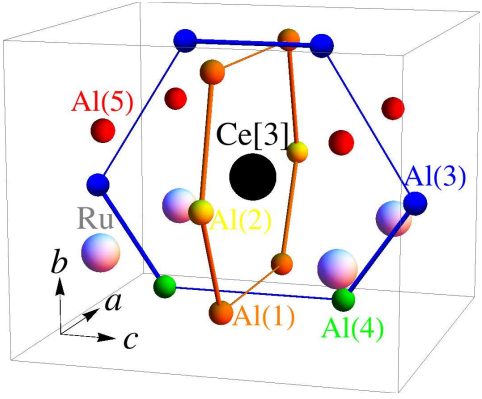}}
\end{center}
\caption{ 
Atomic environment of Ce, surrounded by neighboring 20 atoms of 4 Ru, 4 Al(1), 2 Al(2), 4 Al(3), 2 Al(4), and 4 Al(5).
}
\label{f2}
\end{figure}

\section{Consideration of $^{27}$Al-NQR Spectra}

Lattice distortion to stabilize the spin-Peierls state\cite{Hanzawa}
must be compatible with the $^{27}$Al-NQR spectra,\cite{Matsumura} the result of which is summarized as follows.
The NQR frequencies $\nu _Q$ (with asymmetry parameter $\eta$) observed above $T_0$ are assigned to each Al site as:
$\nu _Q =$ 1.66 MHz ($\eta =$ 0.58) for Al(1), 1.39 MHz ($\sim 0$) for Al(3), 1.77 MHz ($\sim 0$) for Al(4), 2.31 MHz ($\sim 0$) for Al(5), and less than 1 MHz for Al(2) (not observed).
As the sample is cooled below $T_0$,
the NQR peaks of these $\nu _Q$'s simply split into two peaks with almost 1 : 1 intensity ratio, except for Al(5).
Not only the NQR peak of Al(5) does not split below $T_0$, but also the frequency shift is negligible within the experimental accuracy,
the fact of which imposes a strong constraint on a possible lattice distortion.

The main origin of the electric field gradient (EFG) at $^{27}$Al nuclei is considered to be
the charge imbalance of the on-site 3{\it p} electrons, giving rise to
$\nu_Q =(3e^2Q/25h) \langle r^{-3} \rangle_{3p} {\varDelta}n_{3p}$ for $I=5/2$,
with ${\varDelta}n_{3p}$ being the charge imbalance.
We take the principal axis of EFG along the $Z$-axis as usual, then we have
$\varDelta n_{3p} = n_{3p_Z}- (n_{3p_X} +n_{3p_Y})/2$ and
$\eta = (3/2) |n_{3p_X} - n_{3p_Y}|/\varDelta n_{3p}$.
The nuclear quadrupole moment of $^{27}$Al is $Q=0.150$~barn\cite{Raghavan}, and
the value of $\langle {r^{-3}} \rangle _{3p}$ has been calculated by the Hartree-Fock approximation as
$\langle {r^{-3}} \rangle _{3p}$=1.0884~a.u.\cite{Fraga}
Using these values, we obtain $\nu_Q =4.603 \times {\mit\Delta}n_{3p}$~MHz.
Therefore, the experimental value $\nu_Q $ ($\eta$), for example 2.31~MHz ($\sim 0$) for Al(5),
can be accounted for if ${\mit\Delta}n_{3p} = 0.502$ and $n_{3p_X} \sim n_{3p_Y}$.

The charge imbalance of ${\mit\Delta}n_{3p} \approx 0.50$ for Al(5) is considerably large for the following reason.
The total number of $3p$ electrons, $n_{3p} \equiv n_{3p_X} + n_{3p_Y} + n_{3p_Z}$, in an Al ion
is considered to be in a range of $1 \lesssim n_{3p} \lesssim 1.3$, and hence
$n_{3p_Z}$ must be about three times larger than $n_{3p_X}$ and $n_{3p_Y}$,
such as $(n_{3p_X}, n_{3p_Y}, n_{3p_Z}) \approx (0.25, 0.25, 0.75)$,
to account for ${\mit\Delta}n_{3p} \approx 0.50$ and $\eta \approx 0$.
The nearest sites to Al(5) are two Ru sites situated at above and below along the $b$-axis
with the distance of 2.579~\AA, which is the shortest among Al-Ru distances.
The angle of these Al(5)-Ru bonds is 171$^\circ$, which is also much different from 
the angles of the other Al-Ru bonds of around 120$^\circ$:
125$^\circ$ (Al(1)), 112$^\circ$ (Al(2)), 120$^\circ$ (Al(3)), and 117$^\circ$ (Al(4)).
These facts suggest that the EFG at Al(5) nuclei originates dominantly from
the hybridization of $3p$ states of Al(5) with $4d$ states of neighboring Ru's. 
Probably, 
their $\sigma$ bonding, ($pd\sigma$), gives rise to the largest $n_{3p_Z}$, with the principal $Z$-axis parallel to the $b$-axis,
whereas their $\pi$ bonding, ($pd\pi$), and the hybridizations with electrons of the other ions contribute to 
$n_{3p_X}$ and $n_{3p_Y}$.
In addition, the EFG at Al(5) shows practically no change below $T_0$ as mentioned before, and hence
we may be able to assume that Ru atoms, as well as Al atoms at Al(5) sites, do not displace below $T_0$.
Furthermore, it is considered that the hybridizations of $3p$ states of Al ions with $4d$ of Ru and also with $5d$ and $4f$ of Ce
are sensitive to the positions of Al relative to those of Ru and Ce, while
the mutual hybridizations between $3p$ states of Al atoms 
possessing essentially the same configurations are not sensitive.
We therefore also assume that small changes of relative positions of Al atoms considered below do not 
alter substantially the amounts of $n_{3p_X}$, $n_{3p_Y}$, and $n_{3p_X}$, and hence the EFG's.

\section{Lattice Distortion in the Spin-Peierls State}

Now, we consider lattice distortion in the spin-Peierls state.\cite{Hanzawa}
For the dimerization of Ce ions, such as between Ce[3] and Ce[4] connected by the cyan bond shown in Fig.~1,
one may usually expect displacements of these Ce ions approaching each other
to gain the exchange energy.
Such displacements, however, may be negligibly small, at least to an extent not to alter the value of $\nu _Q$ at Al(5).
Instead, Al atoms at Al(1), Al(2), Al(3), and Al(4) sites will displace so as to stabilize the spin-Peierls state.
These Al sites to displace are shown in Fig.~2, by connecting orange lines for Al(1) and Al(2), and blue lines for Al(3) and Al(4).
The central Ce site is contained in the planes constructed by these lines, namely an $ab$ plane for Al(1) and Al(2) and a $bc$ plane for Al(3) and Al(4).
Therefore, it is considered that Al atoms at Al(1) and Al(2) sites displace in the $ab$ plane, and
those at Al(3) and Al(4) displace in the $bc$ plane.
The dimerization will occur along the c-axis, so that we must primarily consider
the displacements of Al atoms at Al(3) and Al(4) sites in the $ac$ plane.

\begin{figure}[tb]
\begin{center}
\scalebox{.84}{\includegraphics{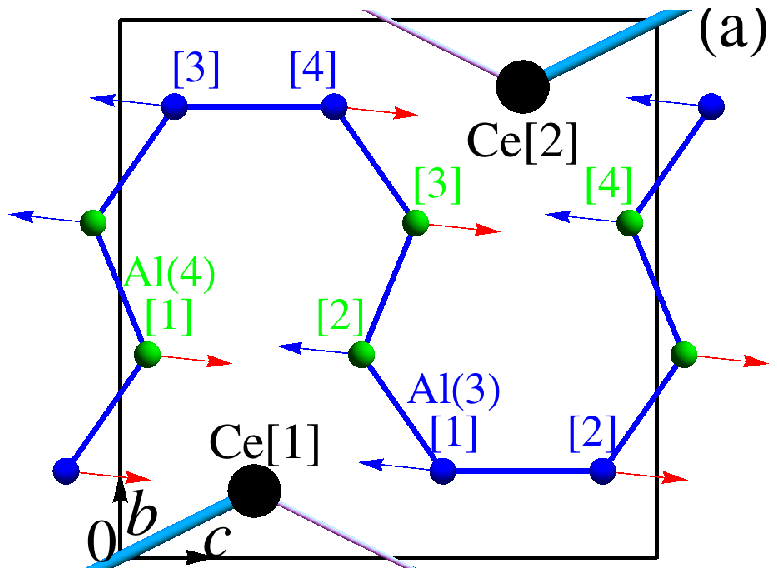}}
\scalebox{.75}{\includegraphics{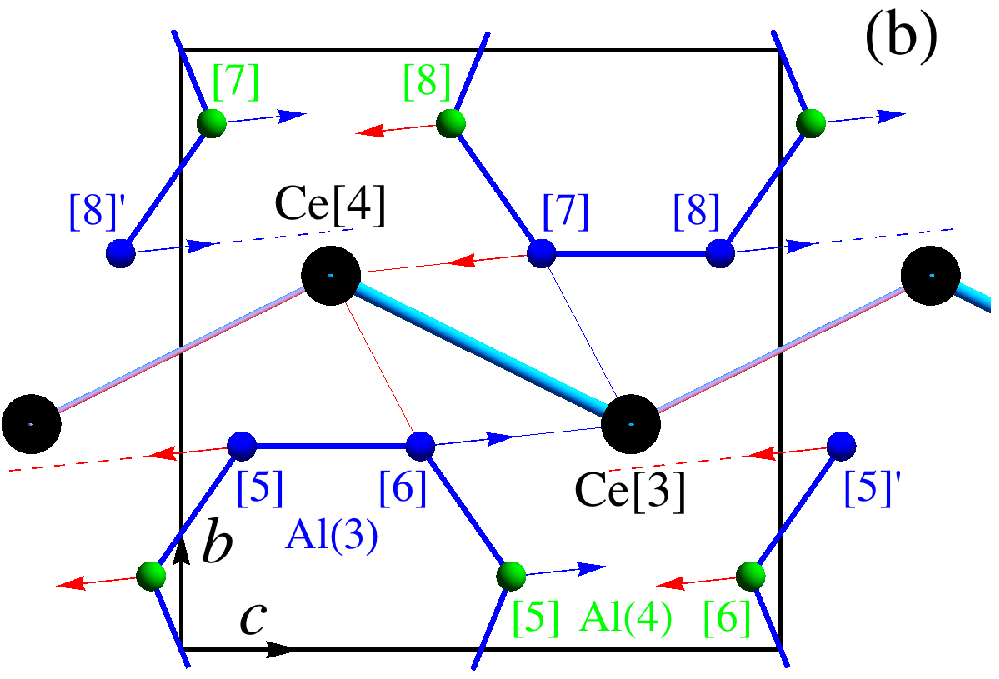}}
\end{center}
\vspace{-0.3cm}
\caption{ 
Displacements of Al atoms in (a) $a=0$ ($x=0$) plane and (b) $a=1/2$ ($x=1/2$) plane, represented by arrows, which are drawn
$10^2$ times larger than the hypothesized displacements described after.
}
\label{f3}
\end{figure}

Figures 3(a) and 3(b) show the atoms in the $a=0$ ($x=0$) and $a=1/2$ ($x=1/2$) $bc$ plane, respectively, where
not only Ce but also Al atoms are numbered from [1] to [8] for the following discussion.
We first discuss the dimerization between Ce[3] and Ce[4] in the $a=1/2$ plane of Fig.~3(b).
If Al atoms approach a neighboring Ce atom, the hybridization matrix elements between the $4f$ states of Ce and
the $3s$ and $3p$ states of Al increases, thereby making the RKKY interaction larger to stabilize
the spin-Peierls state.
As seen from Figs.~3(a) and 3(b),
the movements of Al atoms at Al(3) sites are crucial to the stabilization of spin-Peierls state realized on the zigzag chain along the $c$-axis.
For Ce[3], there exist two n.n. Al(3) sites, [7] and [8], separated by $d \equiv$ 3.230~\AA,
and two n.n.n. sites, [5]' and [6], by $d ^{\prime}\equiv$ 3.249~\AA,
which is only 0.019~\AA (0.59~\%) larger than $d$.
Similarly, Ce[4] has 2 n.n. sites, [5] and [6], and 2 n.n.n. sites, [7] and [8]'.
Instead, an Al(3) site, e.g. Al(3)[6] has Ce[4] as the n.n. site and Ce[3] as the n.n.n. site.

Due to the characteristics of the crystal structure,
an Al atom at Al(3) site is considered to be hard to displace toward its n.n. Ce site, but easy toward its n.n.n. Ce site.
That is, Al at Al(3)[6] will not displace toward Ce[4] but toward Ce[3], as indicated by the blue arrow in Fig.~3(b).
Similarly, Al at Al(3)[7] will displace toward Ce[4] as indicated by the red arrow.
However, if all Al atoms at Al(3) sites displace toward their n.n.n. Ce sites,
e.g. Al at Al(3)[5]' also displaces toward its n.n.n. Ce[3], 
the EFG's of Al(3) will change all together without the observed two-site splitting of $\nu _Q$.\cite{Matsumura}
Instead, Al at Al(3)[5]' will displace in the same way as Al at Al(3)[7] indicated by the red arrow,
which deviates form the direction toward Ce[3].
As a result, Al atoms at Al(3) sites will displace collectively as indicated by the blue and red arrows shown in Fig.~3(b), and
then we obtain the desired two-site splitting of $\nu _Q$ for Al(3).\cite{Matsumura}
Using the atomic coordinates $(x, y,z)=(1/2, 0.3762, 3/4)$ for Ce[3] and $(1/2, 0.3393, 0.3989)$ for Al(3)[6],\cite{Tursina}
the unit vectors parallel to the blue and red arrows are obtained as $\pm (0, \sin \theta, \cos \theta)=\pm ( 0, 0.1045, 0.9945)$
with respect to atomic coordinates,
where $\theta =\pi/30.0$ [rad] $= 6.00^{\circ}$.
Note that the real angle of the arrows to the $ac$ plane is given by $\tan^{-1}(0.1045 b/0.9945 c) = 6.71^{\circ}$.

\begin{figure}[tb]
\begin{center}
\scalebox{.65}{\includegraphics{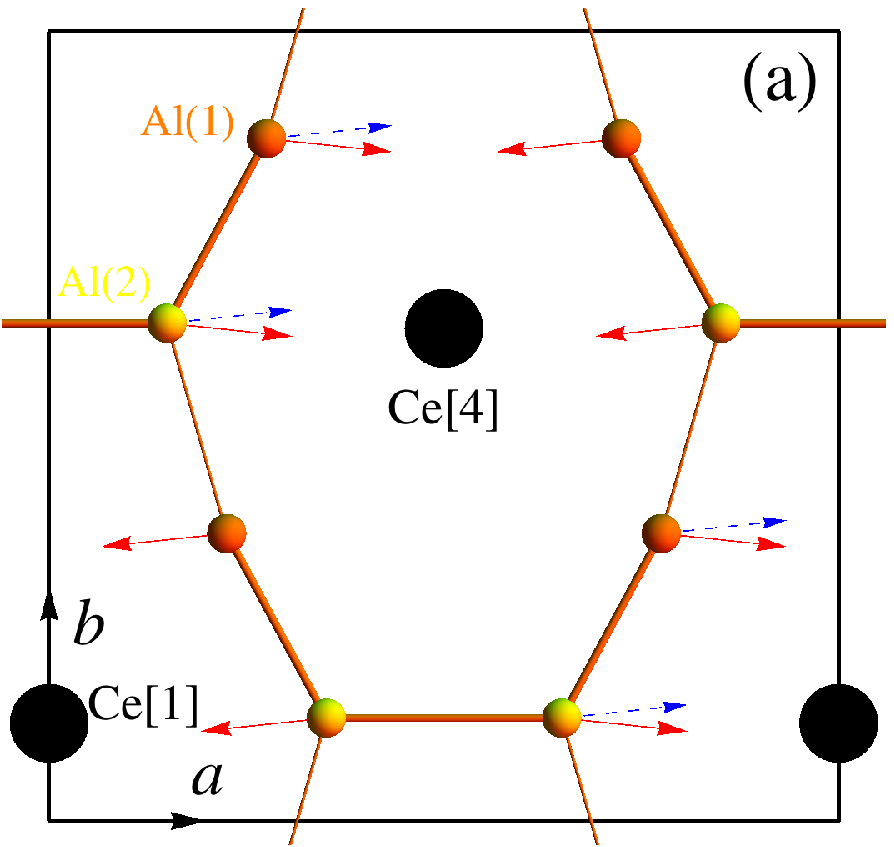}}
\scalebox{.65}{\includegraphics{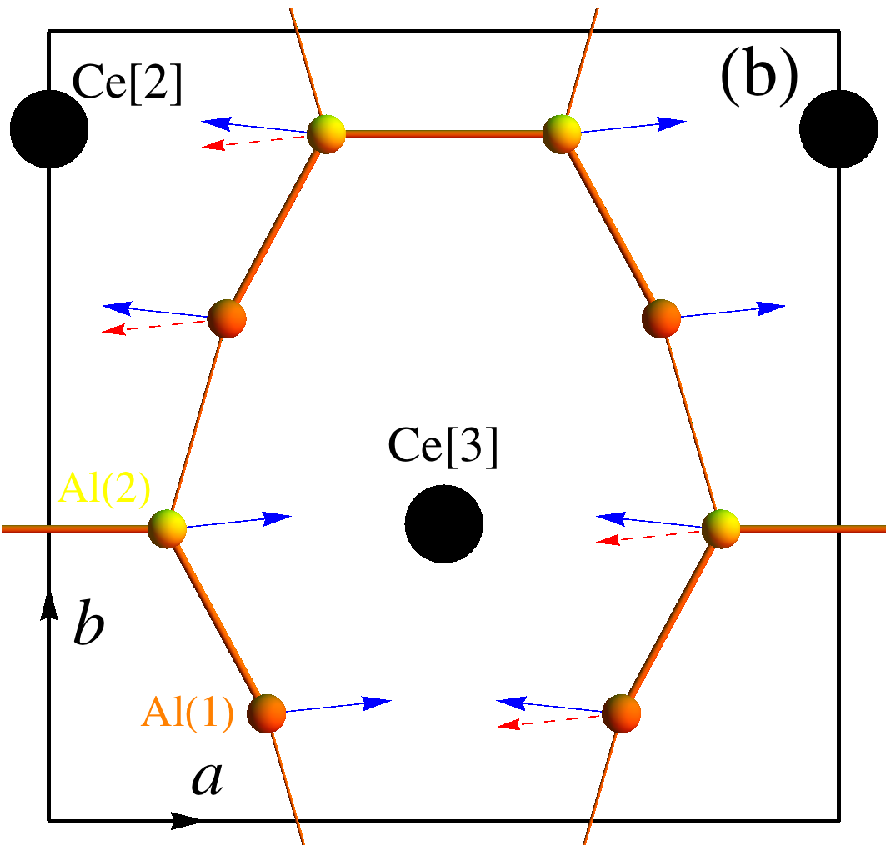}}
\end{center}
\caption{ 
Displacements of Al atoms in (a) $c=1/4$ ($z=1/4$) plane and (b) $c=3/4$ ($z=3/4$) plane.
}
\label{f4}
\end{figure}

The Al atoms at Al(4) sites are considered to displace cooperatively with those at Al(3), as shown in Fig.~3(b),
ensuring the two-site splitting of $\nu _Q$ for Al(4) in the same way as Al(3).
The movements of 
[6] and [7] of Al(3) and [5] and [8] of Al(4) in Fig.~3(b)
will increase the effective interaction to stabilize the dimerization of Ce ions, between Ce[3] and Ce[4],
although such a consideration for the effective interaction
is based on not so much an RKKY mechanism as a superexchange one.
Displacements in the $a=0$ plane relative to those in the $a=1/2$ plane are determined so as to break the invariance of the centering translation
suggested by the neutron diffraction experiments,\cite{Robert} as shown in Fig.~3(a).
The deviations of displacements from the $ac$ plane, indicated by the $b$ components, break the invariance.
The resulting regular lattice distortion will turn
the 1D spin-Peierls order realized individually on each zigzag chain\cite{Hanzawa} into
a three-dimensional (3D) order with the ordering vector $\bm{Q} = (0, 0, 1)2\pi/c$,
as shown in Figs.~3(a) and 3(b), and also in Fig.~1.
Note that the centers of bonds connecting dimerized pair of Ce ions construct a body-centered structure, and
the directions of bonds form an alternating, namely \textit{antiferro} ordering.
Because of $\bm{Q} = (0, 0, 1)2\pi/c$ for a body-centered structure,
$\bm{Q} = (1, 0, 0)2\pi/a$ and $(0, 1, 0)2\pi/b$ are equivalently valid.

Owing to the above lattice distortion of Al(3) and Al(4),
the space group $Cmcm$ (\#63) is reduced to its subgroup $Pmnn$ ($Pnnm$, \#58).\cite{Hahn}
The 8f sites for Al(3) and Al(4) in $Cmcm$ split into two different 4g sites in $Pmnn$, and 
the 8e sites for Al(5) into 4e and 4f sites.
The 4e (4f) sites have two n.n. Ce's belonging to the same dimer (different dimers). Note again that
the EFG at Al(5) has been assumed to be unaffected by small displacements of surrounding Al's.\cite{note}
In $Pmnn$, Al(1) and Al(2) sites are represented by the general points of 8h site,\cite{Hahn}
and hence, as far as the symmetry of $Pmnn$ is preserved,
Al(1) and Al(2) cannot exhibit two-site splitting.
Displacements of Al atoms at Al(1) and Al(2)
may occur 
as shown in Figs.~4(a) and 4(b)
(without dashed arrows), which are determined so as to stabilize the spin-Peierls state by means of decreasing
the distances of Al sites to their n.n. Ce sites to enhance the Heisenberg interaction.\cite{Hanzawa}
To account for the two-site splitting of $\nu _Q$ for Al(1) and Al(2), we should assume a further symmetry reduction
to a subgroup of $Pmnn$
such as $Pmn2_1$ (\#31) and $P2nn$ ($Pnn2$, \#34),\cite{Hahn} although its origin is unclear at present.
For $P2nn$, the displacements of Al's are shown by dashed arrows in Figs.~4(a) and 4(b).


%
In Fig.~5, we show a view of the displacements of Al atoms in the second '$ac$' plane including Ce[3] site, with Ru atoms
connecting the first and second $ac$ planes.
Note that the crystal structure of CeRu$_2$Al$_{10}$ shown in Fig.~1 can be regarded as a sequence of 8 $ac$ layers (planes) as follows:
from the bottom along the $b(y)$-axis,
[Al(5)(on $y=0$)]$-$[1st:Ce[1]-Al(1-4)]$-$[Ru($y=1/4$)]$-$[2nd:Ce[3]-Al(1-4)]$-$[Al(5)($y=1/2$)]$-$[3rd:Ce[4]-Al(1-4)]$-$[Ru($y=3/4$)]$-$[4th:Ce[2]-Al(1-4)]($-$[Al(5)]($y=1$)).
%
%

\begin{figure}[tb]
\begin{center}
\scalebox{.7}{\includegraphics{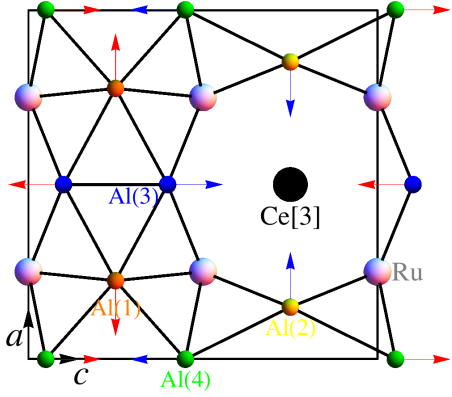}}
\end{center}
\vspace{-0.3cm}
\caption{ 
Displacements of Al atoms in the second '$ac$' plane, with Ru atoms connecting this plane and
the first '$ac$' plane.
}
\label{f5}
\end{figure}

The qualitative discussion given so far may be somewhat obscure, therefore we make
an quantitative estimate of hypothesized displacements in the lattice distortion.
Here, we assume that the Al atom at Al(3) site approaches its n.n.n. Ce sites to the position where
the distance $d^{\prime}$ to its n.n.n. Ce site becomes equal to that $d$ to its n.n. Ce site.
Such a situation is obtained provided that the displacement is given by
$(\varDelta x, \varDelta y, \varDelta z) = 0.0016 (0, \pm 0.1045, \pm 0.9945) = (0, \pm 0.00017, \pm 0.00159)$ in atomic coordinates,
which is equal to (0, $\pm$0.0017 \AA, $\pm$0.0146 \AA), thereby
changing $(d, d^{\prime})=$(3.230~\AA, 3.249~\AA) into (3.234~\AA, 3.234~\AA).
Also for the displacements of other Al sites, indicated by the blue and red arrows in Figs.~3(a), 3(b), 4(a) and 4(b),
we simply assume the same deformations as Al(3) in atomic coordinates, with interchanging $\varDelta x$ and $\varDelta z$
for Al(1) and Al(2).
In that case, we obtain the variations of distances from Al atoms to neighboring Ce sites, as well as to Ru sites,
as shown in Table~I,
where those in parentheses for Al(1) and Al(2) correspond with the dashed arrows in Figs.~4a and 4b.
Note that the lengths of the blue and red arrows in Figs.~3(a), 3(b), 4(a) and 4(b) have been drawn $10^2$ times larger than
these hypothetical displacements.

\begin{table}[t]
	\caption[eigen]{Distances 
	from Al to Ce ($d_{\textrm{Ce}}$) and Ru ($d_{\textrm{Ru}}$) above $T_0$,
	and those variations ($\varDelta d_{\textrm{Ce}}$, $\varDelta d_{\textrm{Ru}}$) below $T_0$ due to hypothesized displacements of Al atoms.}
	\label{tab1}
	\vspace{0.2cm}
	\begin{center}
	\begin{tabular}{ccccc} \hline
		atom & $d_{\textrm{Ce}}$ [\AA] & $\varDelta d_{\textrm{Ce}}$ [\AA] & $d_{\textrm{Ru}}$ [\AA] & $\varDelta d_{\textrm{Ru}}$ [\AA] \\ \hline
%
		Al(1) & 3.212 & $-0.011$  & 2.590 & +0.001 \\
						~ & ~ & ($-0.008$) & ~ &  (+0.002) \\ \hline
		Al(2) & 3.203 & $-0.014$ & 2.764 & $-0.006$ \\
						~ & ~ &  ($-0.014$)  & ~ & ($-0.004$) \\ \hline
		Al(3) & 3.230 & $+0.004$ & 2.629 & $-0.004$ \\
						~ & 3.249 & $-0.015$ & ~ &  ~ \\
		~Al(3)$^{\prime}$  & 3.230 & $+0.007$  &  2.629  & $-0.005$ \\
						~ & 3.249 & $-0.014$  &  ~  & ~ \\ \hline
		Al(4) & 3.188 & $-0.010$ & 2.672 & $+0.001$ \\
						~ & ~ & $-0.007$  & ~ & $+0.003$ \\ \hline
		Al(5) & 3.349 & 0~~~ & 2.579 & 0~~~ \\ \hline
	\end{tabular}
	\end{center}
	\vspace*{-0.2cm}
\end{table}%
%

For the displacements we proposed,
the Al atoms move along the directions normal to the straight lines connecting their n.n. Ru sites,
namely so as to change the angle of Ru-Al-Ru bonds, as shown in Fig.~5.
The angles of Ru-Al(1)-Ru and Ru-Al(4)-Ru decrease to make the corresponding Al-Ru bond lengths increase,
whereas those of Ru-Al(2)-Ru and Ru-Al(3)-Ru increase to make the bond lengths decrease.
The resulting variations $\varDelta d_{\textrm{Ru}}$ of the Al-Ru bond lengths are not so large of the order of $10^{-3}$~\AA.
On the other hand, the variations $\varDelta d_{\textrm{Ce}}$ of the Al-Ce bond lengths are of one order larger than $\varDelta d_{\textrm{Ru}}$,
although $d_{\textrm{Ru}}$'s are smaller than $d_{\textrm{Ce}}$'s.
Considering that 
the Al(4) site has the shortest $d_{\textrm{Ce}}$ and the largest two-site splitting of $\nu _Q$,\cite{Matsumura}
the variations of $d_{\textrm{Ce}}$ may be most effective in changing the EFG's at Al sites.
Because Al atoms at Al(2) sites approach their n.n. Ce site from both sides separated along the $a$-axis
as shown in Figs.~4(a) and 4(b),
%
the distances $d_{\textrm{Ce}}$ of Al(2) sites change largely by almost the same amount. 
It follows that the corresponding EFG is expected to change substantially but split slightly,
in agreement with the experimental result shown in Fig.~3 of ref.~3 by the yellow lines,
which become observable below $T_0$ but exhibit very small splitting.

\section{Concluding Remarks}

The way of tilting of displacements at Al(3) and Al(4) sites from the $c$ direction, namely the sign of the $b$ components and those arrangement,
plays a key role in forming dimerized pairs of Ce ions with the ordering vector of (0, 0, 1), 
as well as in breaking the invariance under the centering translation.
The magnitude of $b$ component is very small of the order of $10^{-3}$ \AA.
Therefore, if there appears a defect to provide an unpaired Ce ion,
the neighboring Al atoms may flexibly follow so as to recompose a dimerized pair.
It may also hold in La substitution for Ce, for which
the spin-Peierls ordering is expected to be robust with increasing La concentration,
similarly to the experimental results of Ce$_x$La$_{1-x}$Ru$_2$Al$_{10}$.\cite{Tanida}
Furthermore, the lattice is expected to shrink at least along the $a$- and $c$-axes due to the proposed displacements, as seen from Fig.~5,
in agreement with a steep shrinkage along the $a$-axis below $T_0$ in CeRu$_2$Al$_{10}$.\cite{Tanida}

In conclusion, we have proposed a relevant lattice distortion to stabilize the spin-Peierls state in CeRu$_2$Al$_{10}$,
compatible with the $^{27}$Al-NQR spectra\cite{Matsumura} and the neutron diffraction patterns.\cite{Robert}
The predicted displacements of Al atoms should be confirmed by x-ray or neutron diffraction experiments.



\begin{thebibliography}{99}
\bibitem{note} Only a symmetry argument cannot account for no splitting of $\nu _Q$ for Al(5), the fact of which
is also true for $Cm2m$ ($Amm2$, \#38) and $Pmnm$ ($Pmmn$, \#59) proposed in ref.~3.
\end{thebibliography}
\end{document}